# Study of helium diffusion in yttria: a multiscale approach based on the density functional theory and kinetic Monte Carlo, with transmission electron microscopy and thermo-desorption spectroscopy.

Vinicius Oliveira Cavalcanti[1], Jérôme Roques[1*], Denis Horlait[2], Eric Gilabert[2], Guillaume Riant[1], Thomas Colombeau-Bedos[3], Nicolas Clavier[3], Aurélie Gentils[1]

[1] Université Paris-Saclay, CNRS/IN2P3, IJCLab, 91405, Orsay, France

[2] Univ. Bordeaux, CNRS, LP2i, UMR5797, F-33170 Gradignan, France.

[3] ICSM, Institut de Chimie Séparative de Marcoule, UMR5257,CEA,CNRS, ENSCM, Univ. Montpellier, Site de Marcoule—Bât. 426, BP 17171,30207 Bagnols/Cèze, France

* Corresponding author: jerome.roques@ijclab.in2p3.fr

## Abstract (maximum of 300 words)

A known issue for future nuclear reactors is helium accumulation inside the steel structure materials, responsible for structural issues such as embrittlement and cracking. One possible solution is using new types of reinforced steel, such as oxide dispersion strengthened (ODS) steel. It consists of adding oxide nanoparticles to the Fe-based material, especially yttrium oxide (yttria, $Y_2O_3$), improving its properties. Therefore, one first step is understanding the helium diffusion inside this system. Very little is known about helium inside yttria, with most studies being theoretical ones. Based on this context, this work proposes a combined theoretical and experimental multiscale approach to investigate helium diffusion inside yttria.

The theoretical approach starts with the density functional theory, used to model the atomic yttria cell and determine helium insertion sites. The transitions between the sites were described using the NEB method. Kinetic Monte Carlo was then employed to obtain the interstitial diffusion coefficient expression. It showed a limited diffusion at temperatures below 600 K, which may indicate a tendency for He to be blocked in the oxide. Then, the charged vacancies were explored. It showed that the vacancy further reduces helium diffusion. Finally, it was demonstrated that helium spreads across different vacancies and interstitial sites.

The experimental part involved implanting helium ions at 50 keV in samples with nanometric or micrometric grains. Then, the specimens were characterised with transmission electron microscopy (TEM) and thermo-desorption spectroscopy (TDS) techniques. TEM did not evidence detectable bubbles even at the highest studied fluence ($1\times10^{16}$ cm$^{-2}$). The TDS highlighted different mechanisms for helium diffusion and the grain size's role, providing a model for diffusion coefficient calculation based on interstitial diffusion.

## 1. Introduction

Ferritic alloys have valuable properties as candidates for future reactor structural materials because of their ability to cope with harsh conditions of high operating temperature (>650 $^{o}$C), pressure and irradiation [1–4]. However, gas accumulation (He, H) inside steels proves to be a challenge to employing these materials [5,6]. He, for instance, is not soluble in metals [1,7,8] and tends to be trapped in vacancies where it accumulates and forms bubbles, which cause structural damage like swelling and embrittlement at high temperatures [7,9–15].

One possible solution to this problem is the employment of reduced-activation ferritic-martensitic oxide-dispersion-strengthened (RAFM ODS) steels [4,16]. They are known for higher mechanical stability with better creep and tensile strength at room and elevated temperatures [17–23]. They also have high stability and resistance to irradiation, avoiding dislocation motion, grain boundary movement and void formation in iron by acting as annihilation sites for defects [5,16,18,24–28]. This improvement comes from a high density of complex oxide nanoparticles evenly dispersed in the ferritic matrix. The type of oxides is usually yttrium oxide (also named yttria, $Y_2O_3$) and Y-Ti-O compounds such as $Y_2Ti_2O_7$ and $Y_2TiO_5$ [16,24,29] due to their stability at high temperatures and no dissolution under displacement cascade damage [21,24,30,31]. The particles are usually pure crystalline 10-40 nm yttria or smaller 1-5 nm Y-Ti-O particles [5,21,23,24].

The helium diffusion in the Fe-based matrix has been studied previously in our group [32]. Regarding the nanoparticle's role in helium accumulation in ODS steels, it is believed that they could trap He atoms, acting as a nucleation centre for tiny helium bubbles [4,5,27,33] and avoiding the irradiation creep and void swelling [5,28]. This ability comes from the high number of possible trapping sites for helium [16,29], which decreases the average He bubble size, preventing or delaying the swelling [4,18,34]. Several experimental studies have investigated the effects of He/H implantation in these materials in the last few years[35–37]. For example, some transmission electron microscopy (TEM) studies were dedicated to the size and distribution of He bubbles in ODS steels [33,38], and others confirmed that He bubbles were more present on the surface of the nanoparticles instead of the iron matrix [35,38,39]. Some other studies are dedicated to the structural changes of the ODS steels after He ion implantation by using techniques such as TEM, electron backscatter diffraction (EBSD), positron annihilation spectroscopy (PAS), and atom probe tomography (APT) [38,40–43]. Likewise, some theoretical works investigated with DFT the formation of oxide nanoparticles in iron alloys and their properties [44,45]. Others were dedicated to defect formation and the influence of the charge using DFT and molecular dynamics (MD) [46–48]. Finally, the interface between ferritic steel and $Y_2O_3$/Y-Ti-O nanoparticles and their interaction with defects and He was also studied using DFT [5,49–52].

Nevertheless, much less is known about the behaviour and diffusion of helium inside the oxide nanoparticles, especially yttria. There is no experimental investigation of helium bubbles in pure yttria specimens or a model of the He diffusion coefficient. Theoretically, a few articles have investigated the helium insertion energies in yttrium oxide by DFT [5,16,24]. One work investigated the role of neutral vacancies in yttria, its helium insertion energy, and the attraction of Y-O divacancy to form stable vacancy clusters [24]. However, no theoretical diffusion coefficient expression was proposed, nor was an analysis of the influence of charged vacancies on the He stability. An article linking both theoretical and experimental approaches to He diffusion inside yttrium oxide is also lacking. This information is fundamental for assessing the usefulness of yttria nanoparticles dispersion in future reactor materials.

This article tries to complete this gap and characterise the He diffusion in yttria with a multi-approach never used before for this system. It is based on calculations at an atomic and macroscopic scale correlated and completed with TDS and TEM. Firstly, the yttria model is optimised and validated using the density functional theory (DFT) and literature results. Second, the possible helium insertion sites are identified, and their insertion energies are calculated. Then, the pathways between these He positions are investigated using the nudged elastic band (NEB) method and the transition state theory (TST). Finally, the kinetic Monte Carlo (KMC) algorithm was used to simulate the He atom's trajectory at different temperatures to determine its diffusion coefficient. The model is then complexified with the influence of charged and neutral vacancies being considered and the accumulation of He inside them. In parallel, experimentally, "pure" yttria samples with different grain sizes were implanted with helium ions at various fluences and characterised using transmission electron microscopy (TEM) and thermo-desorption spectroscopy (TDS) techniques. From these results, a model of He accommodation and diffusion in $Y_2O_3$ was proposed.

## 2. Materials and methods

### 2.1. Theoretical part

Periodic DFT [53] based on the VASP (Vienna ab-initio simulation program) code developed by the Fakultät für Physik of the Universität Wien [54,55] was employed for the structure optimisation of yttria. The system's total energy was calculated using a plane-wave basis set and the project augmented plane wave (PAW) method. It allows the description of the interactions between valence electrons and the core electrons representation based on pseudopotentials [56,57]. They were described based on three functionals: the local density approximation (LDA) and the generalized gradient approximation (GGA), with the Perdew-Burke-Ernzerhof (PBE) and the Perdew-Wang (PW91) [58–60]. The number of valence electrons for Y, O, and He are, respectively: eleven ($4s^2 4p^6 5s^2 4d^1$), six ($2s^2 2p^4$), and two electrons ($1s^2$). The optimised calculation parameters were 400 eV for the cut-off in energy and 0.2 eV for the Gaussian smearing. The yttria unit cell is cubic with 80 atoms (32 Y and 48 O), belonging to the space group $Ia\bar{3}$, number 206 [61]. This cell is already big enough to accommodate the helium atom inserted without much relaxation. The Brillouin zone (BZ) was sampled by the Monkhorst-Pack scheme[62], with an optimized $k$-point mesh of 3×3×3. The convergence parameters were $10^{-6}$ eV per atom for energy until the Hellmann-Feynman forces were lower than $10^{-5}$ eV/Å. The spin-polarization treatment was employed.

Once the yttria unit cell was optimised, the possible helium insertion sites were identified with their formation energies. Consequently, the migration energies between all potential insertion sites and their minimum energy path (MEP) were calculated using the nudged elastic band (NEB) method [63] based on a series of images added as a chain of states between the initial and final sites [64,65]. The spring constant value used in our calculations was 5 eV/Å$^2$. In all the intermediate states, an optimisation of the forces along the perpendicular path direction is made, resulting in the MEP (minimum energy path). The highest energy point, the saddle point or transition state, is then identified and optimised using the climbing image NEB (CI-NEB), which accounts for a more accurate description [66].

The interstitial helium diffusion inside yttria is a thermally activated process. The transition state theory (TST) provides the basis for this type of diffusion (TST) [67–71] as well as the Vineyard approximation [68]. Previous works from our group show the methodology in more detail [72–75], especially our latest work [32].

The jump probabilities are calculated and then used as input values for the kinetic Monte Carlo (KMC) method [76,77]. It is based on a stochastic approach, extending the simulations' time to model the atomic diffusion based on the correspondent jump probabilities and energies. Both KMC and the TST principles are combined to arrive at the diffusion value. The latter says that the sum of the reciprocal of the jump probabilities equals the residence time, which is the required time to pass from one interstitial site to another. Consequently, the Einstein equation is used to determine the helium diffusion coefficient at a determined temperature ($D_T$) as it relates the time with the mean square displacement ($<r^2(t)>$). This equation can be seen in (Eq.1).

$$D_T = \frac{<r^2(t)>}{2 \times d \times t} \qquad (Eq.1)$$

Where $d$ is the diffusion dimension (three as all the directions inside the cell are considered), and $t$ is the time. Based on the equation, the diffusion coefficient is obtained for different temperatures. Then, the Arrhenius equation [78] is employed to arrive at a general relation between the diffusion coefficient and the temperature, as represented in (Eq.2).

$$D_T = D_0 \, e^{-\frac{E_a}{k_B T}} \qquad (Eq.2)$$

Two fundamental parameters are the $D_0$, known as the pre-exponential factor; and $E_a$ which is the effective activation energy. Both could be obtained from a linear regression of the $\ln D_T$ versus $\frac{1}{k_B T}$ plot.

The calculations with charged vacancies in this work were done by manually setting the number of valence electrons in the system. The total energy calculations in these charged systems were employed using monopole-monopole corrections for more accurate results [79].

### 2.2. Experimental part

This section describes the sample preparation of pure yttrium poly-crystals of larger (several μm) and smaller (several nm) grains' sizes, the $^4$He$^+$ ions implantations, and the post-characterisation by Thermo-Desorption Spectroscopy (TDS) and Transmission Electron Microscopy (TEM). The methodology is also similar to the newest work from the group [32].

#### 2.2.1. Materials and ion implantation

The pure yttrium samples with nanograins are a Neyco ultrapure (99.99%) square material, while the micrograined ultrapure samples (99.99%) were specially fabricated and sintered by the *Institute de Chimie Separative de Marcoule* (ICSM), Bagnols sur Cèze, France. The first are squares of 5x5x3 mm, while the second are discs with 8 mm diameter and 0.5 mm depth. The latter were prepared through the synthesis of yttria nanopowders by adapting the protocol of Clavier et al. [80] based on the precipitation and rapid ageing of metallic hydrolysis and using all chemicals supplied by Merck with analytical grade. Firstly, a Y(III) solution was obtained from Y(NO3)3.6H2O dissolution in $HNO_3$. Then it was poured in excess of 2M $NH_3$ (around 400%) under magnetic stirring, forming a Y(OH)$_3$ precipitate grown into the $Y_2O_3$.nH$_2$O form. After separation by centrifugation, a washing and drying phase, and pressure using a tungsten carbide die, the pellets were sintered at 1700 °C for 10 hours under ambient air.

Before the implantation, all the samples were mechanically polished using conventional silicon carbide abrasive paper of small grain sizes to reduce surface damage. The average size of the grains was measured with TEM as 142 nm for the nanograin samples and measured by optical microscopy as 17 μm for the micrograin ones.

The helium ion implantations happened in the JANNuS-Orsay MOSAIC platform of IJCLab (Orsay, France) [81,82]. A 190 kV ion implanter, IRMA, equipped with a Bernas-Nier source, was used. Helium was incorporated with fluences of $1\times10^{13}$, $1\times10^{14}$, $1\times10^{15}$ and $1\times10^{16}$ He ions/cm$^2$. Before the experiments, the Stopping and Range of Ions in Matter (SRIM-2008) code [83] was used to visualise the implanted He ions' profile in yttria and determine, based on the depth, the implantation energy of 50 keV. Full damage cascades with displacement energy of 57 eV for yttrium and oxygen atoms were considered in the code using a previous result from the literature [84]. Figure 1 shows the SRIM-calculated implantation profile of helium ions as a function of the depth. All the implantations were realised at room temperature, and the mean ion flux was $1\times10^{12}$ cm$^{-2}$s$^{-1}$ for all the experiments. All implantations were under vacuum conditions, and the incident helium ions were positioned perpendicularly to the specimens. As a representative example, the damage profile for the highest fluence was chosen for the plot in Figure 1. The peak He concentration (roughly 280 nm) arrives with a He concentration of around 0.7 %.

### 2.2.2. Transmission electron microscopy (TEM)

Once implanted, the samples were analysed using two different techniques. Conventional TEM is one of them at the JANNuS-Orsay MOSAIC platform [81,82]. We employed a 200 kV FEI Tecnai G$^2$ 20 Twin microscope with LaB$_6$ filament and equipped with a Gatan imaging filter (GIF) tridiem apparatus, with thickness measurements using Electron Energy Loss Spectroscopy (EELS) as previously described in [85]. Before the visualisation, the Focused Ion Beam standard lift-out technique was used to extract cross-section TEM thin foils at IEMN, Lille. EELS was used to measure the thicknesses for each studied zone, which were in the range of 45-60 nm. The conventional cavity imaging technique using through-focal series did not confirm the presence of eventual cavity/He bubbles as a function of the helium ion implantation depth, which may still be present but smaller than our microscope resolution (*i.e.* cavities diameter < 0.5 - 1 nm).

### 2.2.3. Thermo-desorption spectroscopy (TDS)

The second characterisation method is thermo-desorption spectroscopy (TDS). The specimens were transferred to the LP2i Bordeaux, where mass spectrometry is employed to detect rare gases down to ultra-trace levels within the PIAGARA platform (French acronym for Interdisciplinary Platform for Noble Gas Analysis) [86–88]. Hence, it can measure the helium release accurately from the samples as a function of the time and temperature. The platform consists of a mass spectrometer with a calibration system for the analysis (A Micromass 12 from VG modified to perform rare gases measurements) [87]; a heating chamber, where the gas is extracted using a focalised laser heating system [88]; and the separation and purification system to eliminate undesired gaseous species. Several pumps are also used to keep the ultra-high vacuum conditions necessary for this type of experiment (routinely $10^{-12}$ bar) [87]. More details on the facility can be seen in our previous works [32,87,88].

Each sample is laid onto a boron nitride support, the implanted side facing up. The laser provides the heating with a natural Gaussian shape, illuminating the sample and the surroundings to around 15-20 mm diameter to ensure temperature homogeneity. Real-time measurement and control of the specimen temperature are made with pyrometers. Three complementary ones are used in this work: a Raytek MM2MHVF1L (1.6 µm; 450 to 2250 °C; spot size Ø <1 mm), an Optris Csmicro (8 – 14 µm; -50 to 1030 °C; spot size Ø of ~11 mm), and a Sensortherm Metis M322 (1.45 – 1.65 µm; 600 to 2300 °C; spot size Ø of ~0.8 mm). The Optris was chosen for temperatures below 600 °C, and the other two were used together for higher temperatures. For yttria, the emissivities ε at the measurement wavelengths are relatively high and constant at temperatures below roughly 1000 °C [89], the range where helium is desorbed, limiting absolute errors. As the Optris pyrometer has a measurement spot size a bit larger than the sample surface (hence integrating partly the BN support), the Optris temperature record was corrected after the experiment, using room-temperature and the >600°C data from the Metis pyrometer as absolute reference points.

The heating chamber was mostly directly connected to the spectrometer through a purification zone, consisting of devices to trap all gases except He and Ne (charcoal cooled to liquid nitrogen temperature, hot getters), with the same configuration as before in our previous work [32]. Two types of experiences are performed: the samples are heated with a 5 K/min ramp from room temperature, or certain plateaux of temperatures are chosen. In the first case, the evolution of the He release profile is observed. The temperature was increased until no more gas was relaxed after an extensive temperature range. The second type is based on following the He release pattern in a material at a given fluence. Then, certain temperatures are chosen, where the gas released is measured in these temperatures' plateaux for several minutes. The goal is to obtain the $^4$He apparent diffusion kinetics at each discrete temperature and arrive at the activation energies through Arrhenius plots. This case is presented and discussed in more detail in Section 3.2.2.2, with an equation representing it.

Whatever the temperature treatment, in order to eradicate problems such as instabilities of mass spectrometer sensitivity and to determine absolute $^4$He release from the sample, a precisely known amount of a $^4$He+$^3$He reference gas was added before any annealing. The monitoring of $^4$He/$^3$He intensity ratios evolution by mass spectrometry then allowed deriving the quantity of $^4$He released as a function of time and temperature.

## 3. Results and discussion

### 3.1. Theoretical part

#### 3.1.1. Yttria unit cell and validation of calculation parameters

Yttria is part of the space group $Ia\bar{3}$ (No. 206) with a bixbyite cubic bcc structure with eighty atoms: 32 Y occupying the 8b (¼, ¼, ¼) and 24d ($u$, 0, ¼) Wyckoff positions; and 48 O at the 48e positions ($x_0$, $y_0$, $z_0$) [16]. The crystal structure is shown in Figure 2 with the calculated lattice parameter of 10.65 Å. We explored different functionals and found GGA PBE gives accurate results (Table 1).

### 3.1.2. Investigation of He insertion sites

Following the optimisation of the unit cell, the possible helium insertion sites in the yttrium oxide were identified. The lattice parameter of the yttria's unit cell is big enough to accommodate the insertion of helium without significant cell relaxation (around 0.1%). Therefore, further calculations were performed at constant volume. Three types of insertion sites were identified and are visible in Figure 3(a-c): 8b (Figure 3-a), equal to the Wyckoff 8b position located between six Y 24d atoms; 16c (Figure 3-b), equivalent to the Wyckoff position 16c, at the centre of six O 48e atoms; and centre (Figure 3-c), situated in the centre of four Y 8b and two Y 24d atoms. The insertion energy was compared to other literature results in Table 1, validating our model. The more stable site is the 16c, followed by the 8b and the centre.

### 3.1.3. Interstitial He migration

For simplification, the yttria helium insertion sites 8b, 16c and centre were recalled as S1, S2 and S3, respectively. Three possible transitions were identified and are shown in Figure 4: the S1-S2 one at 2.15 Å distance (shown in Figure 4-a), the S1-S3 at 3.71 Å (in Figure 4-b), and the S2-S3 at 2.41 Å (seen in Figure 4-c). The NEB method was employed to calculate the migration energies and identify the transition states. Our results agree with the literature (0.31 to 0.31 eV [24] for the transition S1-S2, and 0.80 to 0.78 eV for the S2-S3 [24]), validating our simulations. Considering these transitions, the path to the helium diffusion inside yttria can be seen in Figure 5-a,b and c. The most probable transition is between S1 to S2 with an energy of 0.17 eV and 0.31 eV in the other direction. This may come as both sites are close and the surrounding atoms are relatively far from the transition state (2.17 Å to an oxygen atom and 2.37 Å to a yttrium one). For the transition from S2 to S3, the migration energy is slightly bigger of 0.37 eV and 0.80 eV from S3 to S2. The sites are a bit further away, and the surrounding atoms are closer (2.00 Å to the oxygen and 2.29 Å to the yttrium atoms). Finally, the transition between S1 and S3 is less probable due to the high transition energy values of 2.35 eV and 2.64 eV: Both sites are far from each other with a distance of 3.71 Å and the surrounding atoms are close to the transition state (1.04 Å for the oxygem and 1.59 Å for the yttrium atoms).

### 3.1.4. Frequency and Kinetic Monte Carlo calculations

With the migration energies, the Vineyard theory[68] was used to calculate the jumping probability expressions for each transition. The vibration frequencies at the stable and transition sites were calculated to obtain the attempt frequencies based on the Vineyard theory. The result for the different transitions is seen in the following equations:

$$\Gamma_{S1-S2} = 5.53 \times 10^{12} exp\left(-\frac{0.17}{k_B T}\right) \quad \text{(Eq.3)}$$

$$\Gamma_{S2-S1} = 5.71 \times 10^{12} exp\left(-\frac{0.31}{k_B T}\right) \quad \text{(Eq.4)}$$

$$\Gamma_{S1-S3} = 3.68 \times 10^{12} exp\left(-\frac{2.64}{k_B T}\right) \quad \text{(Eq.5)}$$

$$\Gamma_{S3-S1} = 7.51 \times 10^{12} exp\left(-\frac{2.35}{k_B T}\right) \quad \text{(Eq.6)}$$

$$\Gamma_{S2-S3} = 4.19 \times 10^{12} exp\left(-\frac{0.80}{k_B T}\right) \quad \text{(Eq.7)}$$

$$\Gamma_{S3-S2} = 8.82 \times 10^{12} exp\left(-\frac{0.37}{k_B T}\right) \quad \text{(Eq.8)}$$

The calculated jumping frequencies were used as input for our KMC homemade code. It allows us to calculate the 3D He trajectory at different temperatures (200,000 steps were used for each simulation).

The trajectory starts to become significant and Brownian towards 800 K. Therefore, helium should not diffuse significantly at low temperatures.

The He trajectories from the KMC code were then used to calculate the diffusion coefficient at different temperatures (300 to 3000 K) based on the mean square displacements. Using the Arrhenius law (Eq.2), the logarithm of the calculated diffusion coefficients $D_T$ and the inverse of the temperature were plotted. It allowed retrieving the pre-exponential factor $D_{0,i}$ and the effective activation energy ($E_a, i$) for the He interstitial diffusion in yttria. The resulting equation for the diffusion is in (Eq.9), and the corresponding curve is visible in a further section of this work (Figure 11).

$$D_i = 1.3 \times 10^{-3} e^{-\frac{0.70}{kT}} \, cm^2 s^{-1} \tag{Eq.9}$$

The activation energy equals 0.70 eV, and the pre-exponential factor is $1.3 \times 10^{-3} \, cm^2 s^{-1}$. These values suggest a lower interstitial diffusion for yttria than pure iron, which will be the central element of the steel for future reactors. Our group provided the iron results in a recent work [32]: 0.06 eV for the activation energy and $6.5 \times 10^{-4} \, cm^2 \cdot s^{-1}$ for the pre-exponential factor. For example, the diffusion value was calculated around 1000 K, the current highest previewed temperature for these materials in fusion reactors[90]. The value found was $3.71 \times 10^{-7} \, cm^2 s^{-1}$, $10^3$ lower than the one for pure iron of $3.23 \times 10^{-4} \, cm^2 s^{-1}$. These results may be an indication of the lower He diffusion inside the oxide in comparison to the metal.

### 3.1.5. Vacancy effect

With the first result of interstitial diffusion, the role of vacancies was also investigated to complexify the model. The particular difficulty is the charged effect coming from the vacancies. Hence, an analysis was realised considering both charged and neutral vacancies: $Y^{3+}$ and $O^{2-}$. The methodology is based on removing either an oxygen or a yttrium atom from the 80-atom yttria unit cell. Three types of vacancies are present: $Y_1$ by removing 8b yttrium atoms, Y2 by removing 24d yttrium atoms, and O by removing an oxygen atom. Then, a helium atom is inserted inside it, and the insertion energies are calculated based on the following equation:

$$E_{sub}^{f} = E_{yttria+He} - (E_{yttria-v} + E_{He})  \qquad (Eq.10)$$

Where $E_{yttria+He}$ is the energy of the cell with helium in a substitutional position, $E_{yttria-v}$ is the energy of a yttria cell with a vacancy and without the He atom, and $E_{He}$ is the energy of an isolated helium atom. Both neutral and charged vacancies were considered. The calculated energies are compared to the literature results, as shown in Table 2.

It is visible that a change in the stability order arrives from the choice of neutral and charged vacancies. In the first case, the Y2 is more stable than the Y1, and this tendency is inverted when considering the charges. This phenomenon could arrive from how the charge is distributed in the two sites, as they have different environments with distinct atoms. The charge distribution plays a role in the stability; therefore, Y1 has a charge distribution that reduces the repulsion with the electronic cloud from helium better than Y2.

Our results differ from the other reference found in the literature [24]. This difference probably comes from the different calculation parameters used. Lai et al. used an energy convergence of 0.01 eV, while we used $1 \times 10^{-6}$ eV. To check these differences, we used similar parameters, and we arrived at similar results with 0.193 (0.180 [24]), 0.213 (0.234 [24] and 2.44 (2.627 [24]) eV for Y1, Y2 and O, respectively.

The next step would be the possible migration pathways considering the vacancy. Various possibilities are likely since there are three types of interstitial sites and three types of vacancies. A diffusion model considering all these possibilities is highly complex and surpasses the scope of this article. Therefore, we investigate the possible paths for the helium to leave the less unstable vacancy (Y1 charged vacancy site) towards an interstitial site. As the vacancy has an attraction effect, the two nearest interstitial sites are unstable as He returns to the centre of the vacancy (Eact ≈ 0 eV). He must be located in the third interstitial site to overcome this attraction, which involves at least 4.45 Å distances (S2 site). Hence, we expect these transitions to be highly unlikely. Another point is the migration energy of at least 1.39 eV, visible in Figure 6, showing the difficulty of leaving the vacancy. However, this value is significantly lower than the iron one, the main element of future nuclear steels, calculated by our group of 2.35 eV [32]. They may not trap He inside as firmly as in Fe.

It is visible that the migration energies will be higher in a model with vacancies, further decreasing helium's diffusion coefficient inside yttria. It could be difficult for the helium to diffuse and accumulate at one specific point. Lai et al. [24] showed that the binding energy of helium to vacancy does not allow the formation of vacancy clusters with many atoms (only up to 6 He), avoiding the formation of helium bubbles in the vacancies. Sun et al. [5] showed that a bulk-perfect unit cell could accumulate many He atoms (up to 50) without restructuration as many different interstitial sites are present. However, details about this distribution (where and how the atoms are organised) were not given. Therefore, we also investigated this phenomenon by progressively adding He atoms (up to 15 atoms) to different interstitial and vacancy sites. The He atoms tend to distribute themselves around the available sites and not accumulate in one site (maximum around six atoms in the case of a vacancy). This phenomenon was also investigated by ab-initio molecular dynamics simulations (as implemented in VASP [54,56] and in a similar method as done in our previous work [32]). It showed the helium atoms distributing themselves around the interstitial sites nearby at temperatures of 800 K and not in one single point after several ps. Therefore, it is clear the role of this oxide as a source of helium accumulation as helium is trapped with a low diffusion coefficient (in comparison to the iron matrix), and there is much space for the gas to be allocated as it has many interstitial sites. This helium should be distributed in different sites and not localised in a specific one (such as a vacancy).

### 3.2. Experimental Part

#### 3.2.1. Transmission electron microscopy (TEM)

Underfocus and overfocus bright-field (BF) TEM images were taken on thin foils extracted from both nanometric and micrometric yttria grain samples implanted with He ions at a fluence of $1\times10^{16}$ at room temperature. Our microscope did not confirm the presence of detectable helium bubbles (diameter larger than 0.5 nm), as visible in Figure 7, an example of micrometric-grained samples. This fact does not ensure that there are no helium bubbles in this material, but that at these conditions of temperature and fluence, they are not big enough to be detected by our TEM. This agrees with our calculations in the previous sections, which showed that He does not accumulate in a vacancy. It settles well with Lai et al. [24], who showed that the vacancies would not allow the accumulation of many atoms, therefore not forming bubbles.

### 3.2.2. Thermo-desorption spectroscopy (TDS)
#### 3.2.2.1. He desorption profiles

After the end of the TDS experiences, the sample surfaces melted, and the experimental implanted fluences were thus determined by the cumulated He releases measured by mass spectrometry. The results for the nanometric grain specimens were: $1.015 \times 10^{13}$, $0.738 \times 10^{14}$, $0.784 \times 10^{15}$ and $0.568 \times 10^{16}$ at. cm$^{-2}$. The equivalent for the micrometric grain specimens are $0.98 \times 10^{14}$, $0.88 \times 10^{15}$ and $0.98 \times 10^{16}$ cm$^{-2}$. These values have an associated error of ±5%. The differences between the targeted and real measured fluences are probably due to one or several factors: it could be due to a saturation of He in the material, possibly coming from blistering with the accumulation of implantation damages and saturation of the specimen [91,92]; the nanometric grains could be pathways to the desorption of the helium, explaining the reason for a bigger difference in this type of sample; other factors could be that the attaches used to hold the samples covered varying areas of the surface or the uncertainties related to the flux measurement (approximately 1% uncertainty for the ion beam [93]); another possible factor is that Rutherford backscattering could become significant with the increase of the fluence; finally, TDS determination of the fluence could have been at its lower bound.

The heating ramp used in the experiences was 5 °C/min from room temperature to around 800°C. The helium desorption profile for each fluence versus the temperature for both nanometric and micrometric grain yttria samples can be seen in Figure 8. Two peak regions are identified in all cases, with the values varying from each fluence: region I from 150-500 °C (in the case of nanometric grains) and from 300-440 °C (in the case of micrometric grains), and region II from 400°C onwards for both nanometric and micrometric grains. Region I comprises most of the He budget (80-95%).

A higher fluence tends to delay the first helium release peak to higher temperatures. For the nanometric first peak $1 \times 10^{13}$ at. cm$^{-2}$, the release arrives at the maximum around 300 °C, while for $1 \times 10^{14}$ at. cm$^{-2}$, it is slightly shifted towards 350 °C (Figure 8-a). In the case of $1 \times 10^{15}$ and $1 \times 10^{16}$ cm$^{-2}$, the maximum releases are closer to one another (450 and 425 °C, respectively), suggesting a saturation effect. This progressive temperature may stem from a change in the mechanism of helium diffusion related to fluence and/or to a growing influence of irradiation defects. The diffusion could be associated with interstitial helium diffusion at lower fluences, which requires lower energy. At higher fluencies, more vacancies are produced in the material. These could be trapping sites for helium with the formation of $He_n$-$V_m$ clusters (where n and m are the number of helium atoms and vacancies, respectively), which require higher energies (thus higher temperatures) for diffusion. In addition, a saturation effect seems to be present around the highest fluence, as there is a slight shift towards lower temperatures for the fluence of $1 \times 10^{16}$ compared to $1 \times 10^{15}$ cm$^{-2}$. It could be due to a saturation of the vacancies and the formation of $He_n$-$V_m$ clusters with higher values of n and m, which may be less stable. It agrees with our theoretical calculations, which showed that the yttrium vacancy could not accumulate many He atoms, and once a limit is passed, the helium atoms go to other sites. The other peak of Region II consists of small peaks and is responsible for less than 20% of the helium desorption. It is seen by slow desorption from 380-450 °C until 700-800 °C. It may be the diffusion of bubbles or stable complexes.

The comparison between nanometric and micrometric grains is visible in Figure 8-c) and d). The same two peak regions and the saturation with $1 \times 10^{16}$ are visible. The second peak region is similar to the case of nanometric grains with a slow diffusion of the remaining helium, probably coming from the bubbles' diffusion. The most significant change is the shift of the first peak towards higher temperatures, from 150-200 °C to 325-350 °C. The peaks are also sharper than the nanometric ones, especially at $1 \times 10^{14}$ and $1 \times 10^{15}$ cm$^{-2}$. Two phenomena could explain such difference: first, we have to account that although most of the He will be implanted 100-400 nm below the sample surface, in nanometric samples, the grain average size is only 142 nm, thus many He will actually be closer to a grain surface. Since grain boundaries are fast-diffusing, a significant part of the He population may have the easiest and shortest way to escape the material, explaining the earliest detection of He by TDS in nanometric grain samples compared to micrometric grains samples. The similar activation energies of He diffusion for nanometric and micrometric grains (later determined) comfort this hypothesis. The second, non-mutually exclusive possibility is that the outer region of the grains allows faster diffusion, possibly through enhanced defect annihilation near grain boundaries [94–96]. Considering, as shown by our theoretical investigation, that defects tend to retain or slow He diffusion, their enhanced annihilation near grain boundaries would have a more pronounced effect in nanometric grain samples than in micrometric grains samples.

In addition to the ramp experiments, three heat treatments with constant temperature plateaux were performed: two with nanometric yttria grains with a fluence of $1 \times 10^{14}$ and $1 \times 10^{15}$ cm$^{-2}$, and one with micrometric grains and a fluence of $1 \times 10^{14}$ cm$^{-2}$. The results are in Figure 9. Based on these results, a model was developed for the diffusion coefficient calculation.

### 3.2.2.2. Helium desorption model

Both Figure 8 and Figure 9 show the difficulty of developing the model with different peaks and diffusion mechanisms for helium diffusion. Based on the literature and theoretical results, some simplifications may be adopted in the model:

- Our theoretical results show that the He interstitial diffusion is around 0.7 eV, with some

transitions available at lower temperatures and others at higher ones. It means that He is trapped at room temperature in these sites, according to our simulations. Therefore, not a significant amount of He should escape before the experiments, at least in the micrometric samples.

- Not much is described about He trapping sites in the literature. It is believed that He could be trapped by defects such as interstitial states, grain boundaries, impurities, vacancies, and helium/vacancy/self-interstitial complexes [5,16,24]. Some may be naturally present, such as grain boundaries, especially in the nanometric samples, while others are induced by the He irradiation, such as vacancies and dislocation and their complexes [24,46,97].

The construction of a model considering all these parameters would require highly complex modelling and computational power work. As a first step, we propose some assumptions to establish a simplified model:

- No surface evaporation is considered (Yttria does not sublimate at ultra-low pressures and temperatures up to 800 °C [98,99]).
- All He concentration at the initial state is trapped, probably into a weak trap, such as interstitial states, where He is easily desorbed with annealing or trapped into a more stable trap (such as vacancies).
- The grain boundaries are not considered for the simulation, and TRIM provides the geometric profile.
- The particularities of each defect, such as the various dislocations and He-V clusters, are not considered and are grouped into generic groups.
- Only one-dimensional diffusion is considered (depth).
- As He leaves a trap, we assumed it diffused quasi-instantaneously towards the surface. However, on average and in reality, the atom is retrapped in other defects several times before arriving at the surface. The probability gets higher with higher fluences as there is more damage to the matrix and, thus, a greater concentration of the diffusion-limiting traps. We consider this possibility by calculating an apparent detrapping constant, which is lower than the real one (inaccessible by the experiments). This determined constant integers for all the successive trapping-detrapping events. However, the mean number of these processes cannot be deduced and known. This consideration is based on our DFT calculation for the interstitial diffusion and previous models developed for He in $UO_2$ and $B_4C$ [100–103] and iron [32].

Taking these aspects, we define the starting parameters for the model using the $B_4C$ model [100] and for iron [32], previously done by our group, as a basis:

- $u$: the first state of He atoms, available to diffuse at the peak I in the ramp experiments, probably stable interstitial sites.
- $v$: second state of He atoms, available to diffuse at peak II in the ramp experiments, which requires more energy to escape yttria, probably related to vacancies and/or more stable interstitial sites.
- $C_u$: Concentration of the first diffusing He atoms (at.m$^{-3}$).
- $C_v$: Second trapped He population concentration.
- $k_{bu}$ and $k_{bv}$: rate of helium atoms that definitively can escape the first and second trap types in the material (s$^{-1}$), respectively.
- $\tau_u$ and $\tau_v$ : time constant for a first-order linear response to the rate of He atoms escaping first and second trap types (s$^{-1}$).
- $D_u$: First available He atoms diffusion coefficient (m$^2$.s$^{-1}$).
- $D_v$: Second trapped He atoms diffusion coefficient (m$^2$.s$^{-1}$).
- $F$: Fraction of atoms available in the first state $u$. Consequently, $(1- F)$ is the fraction of atoms initially available in the second state $v$. For the Yttria 1×10$^{14}$ Micro model, $F$ is considered equal to 0.85. For both Yttria 1×10$^{14}$ and 1×10$^{15}$ Nano models, $F$ is considered equivalent to 0.90 on the basis of the release profile in Figure 8-c.

Based on Fick's laws, the following equations are used for each temperature plateau as a function of time $t$ and depth x, as seen in (Eq.11):

$$\begin{cases} \frac{\partial C_u}{\partial t} = D_u \frac{\partial^2 C_u}{\partial x^2} - k_{bu} * e^{-\frac{t}{\tau_u}} * C_u \\ \frac{\partial C_v}{\partial t} = D_v \frac{\partial^2 C_v}{\partial x^2} - k_{bv} * e^{-\frac{t}{\tau_v}} * C_v \end{cases} \qquad (Eq.11)$$

For all these equations' systems, boundary conditions are necessary. They consist of zero concentrations at the surfaces at any time:

$$C_u(0,t) = C_v(0,t) = C_u(e,t) = C_v(e,t) = 0 \qquad (Eq.12)$$

The TRIM calculations provide the initial free He profile in the models [83] (Figure 1). Then, the following can be written:

$$\begin{aligned} C_u(x, t=0) &= F * C_{TRIM}(x), \text{ and} \\ C_v(x, t=0) &= (1-F) * C_{TRIM}(x) \end{aligned} \qquad (Eq.13)$$

The differential equations were resolved by finite element analysis employing the FlexPDE software (PDE Solutions, Inc.)[104]. The total gas released content $R_t$ is calculated as:

$$R_t(t) = 1 - \frac{\int_0^\infty C_u(x,t)\,dx + \int_0^\infty C_v(x,t)\,dx}{\int_0^\infty C_{TRIM}(x)\,dx} \qquad (Eq.14)$$

The iteration loop used to fit the data was designed in-house based on the module kmpfit from the Python Kapteyn library [105]. Each parameter gets an error value assigned calculated from the standard deviation of the fit and experimental error bars. The equations for each temperature plateau shown in

Figure 9 were resolved based on the previously described models. The initial states $C_u$ and $C_v$ are considered the final states of the previous step. At the end of the iteration loop, the determined parameters $(D_u, D_v, k_{bu}, k_{bv}, \tau_u, \tau_v)$ were used to reproduce the ramp release results shown in Figure 8. An example is visible in Figure 10 for the $1\times10^{14}$ micrograined specimen. The fitted activation energies were varied by only 15% maximum to retrieve these results. The excellent correspondence between these two curves and the small error (Figure 9-a) show that this model correctly and efficiently reproduces the experimental data.

Based on the previous equations, the diffusion coefficient at different temperatures was calculated for each sample in Figure 9. A plot comparing our DFT results (section 3.1.4) and the TDS results is seen in Figure 11, complemented by Table 3.

Firstly, in Figure 11, it is visible that the TDS results have lower diffusion coefficients than the DFT ones by several orders of magnitudes. It comes from the fact that our DFT calculations for diffusion did not consider defects, such as vacancies. Comparing the TDS results, both nanograined samples have higher diffusion coefficients at lower temperatures (~450-700 K) than the micrograined specimen. This is in agreement with the observations in Figure 8-c), as the nanograined samples start quantitatively releasing He at a lower temperature than the micrograined ones, and as discussed in the previous section, this should be related indirectly or directly to the higher concentration of grain boundaries. Then, around 700-800 K, all samples have a similar diffusion coefficient. It may come from a different mechanism with the formation of more stable vacancy clusters due to the higher fluencies, which will reduce diffusion. The results in this area also have a higher error bar associated, and there are only one or two measurement points by sample, which makes it challenging to conclude about the mechanisms.

Table 3 shows the different regions of the activation energy and pre-exponential factor. For all three specimens, there is the first region with low values, where the pre-exponential factor $D_0$ is around $10^{-11}$ m$^2$s$^{-1}$, and the apparent activation energy of diffusion $E_a$ is about 0.7 to 0.9 eV. It is around the same low-temperature (450-673 K) region seen in the last paragraph. More importantly, it is notorious the similarity between the activation energy calculated from DFT (0.70 eV) and the experimentally measured ones (0.777, 0.730 and 0.871 eV for the $1\times10^{14}$ nanograined, $1\times10^{15}$ nanograined and $1\times10^{14}$ micrograined samples, respectively). It suggests that the active diffusion mechanism is the interstitial diffusion accounted for in DFT. The differences between the experimental and theoretical results in Figure 11, as the first are lower than the second, most likely come from the pre-exponential factor, also called the frequency factor. It may arrive that the DFT calculations overestimate this value or do not consider possible traps (such as vacancies or dislocations), which do not forbid diffusion nor add a high cost in terms of energy but only temporarily halt He diffusion, reducing the $D_0$ value. Nevertheless, the interstitial diffusion mechanism is likely the most prominent in this region, as the $E_a$ values by DFT and TDS are very similar.

It is also visible in Table 3, a second region (573-668 K), with higher pre-exponential factors and activation energies, only for the $1\times10^{14}$ specimens (both nano- and micrograined) as there were no experiments done at these temperatures for the $1\times10^{15}$ sample. The pre-exponential factor is around $10^{18}$ times higher in the micrograined sample and $10^{26}$ in the nanograined one. This is visible in the slope change of these specimens in Figure 11. The activation energy is significantly increased to 2.9 eV in the micrograined and 4.0 eV in the nanograined samples. This may be related to an extended defect, like dislocations, or because of closed porosity, which is associated with the size of the grains. In any case, there are few points in this region, making it harder to have a definitive conclusion. Further studies should be done considering more prolonged plateaux and reduced temperature steps. It would allow a more complex model to be developed, considering the various mechanisms of He diffusion in yttria.

## 4. Conclusions

Combining DFT and kinetic Monte Carlo approaches allowed us to identify the three possible helium insertion sites and diverse possible transitions. Lower interstitial He diffusion than in iron was calculated for a perfect cell. The vacancy's role was also investigated, and the charge's consideration changed the insertion energy. Moreover, one possible transition to leave from a Y1 vacancy was examined, showing a high migration energy of 1.39 eV. Using DFT and BOMD, several helium atoms were added to the vacancy and the interstitial sites. He seems not to accumulate or grow in yttria (no self-interstitial atom formation as observed previously in iron).

The second part used TEM and TDS experimental techniques to characterise yttria samples with nanometric and micrometric grains after helium implantation. No bubbles were observed by TEM at the highest point of the implantation profile or at the grain boundaries at our maximum He implantation fluence of $1\times10^{16}$ cm$^{-2}$. It does not indicate that He was not there, but the bubbles could be too small to be observable with our microscope (i.e. below 0.5 nm in diameter). This observation is in agreement with our DFT calculations as He should not accumulate so quickly inside the oxide.

Finally, the TDS complemented these results by showing the desorption profile as a function of temperature. Higher implanted fluences showed a slightly delayed diffusion, probably due to the formation of some traps, such as helium-vacancy clusters, due to the implantation-induced damage to the matrix. Two zones were identified, a first peak related to He initially residing in (less stable) interstitial states or defects and another with an efficient trapping capability for He, only releasing it over about 500 °C. The nanograined samples showed the same mechanism but with an early desorption, likely associated with grain boundaries diffusion. Lastly, the helium diffusion coefficient expression was modelled based on three TDS plateaux. They corresponded very well with the experimental data and allowed a comparison with our theoretical results from DFT. The similar activation energy results between DFT and TDS for the region at lower temperatures show a good correspondence between both methods and confirm that the diffusion in this region is likely mainly driven by the interstitial mechanism.

The results emphasize the helium diffusion in yttria, with its distribution even with vacancies. The effect of yttria nanoparticles dispersed in an iron matrix, especially at the interface between both systems, will be the next step to study.

## 5. Acknowledgements

One author (V.O.C.) is grateful for funding from the doctoral school PHENIICS (Université Paris-Saclay). We also appreciate Christophe Diarra and Emmanouil Vamvakopoulos, who helped us manage the calculations cluster: GRIF (http://www.grif.fr). We want to acknowledge Florian Pallier for the help in the preparation of the samples, as well as the technical staff of the JANNuS-Orsay MOSAIC platform at IJCLab, and in particular, Jérôme Bourçois, Silvin Hervé, Philippe Benoit-Lamaitrie and Cédric Baumier. The French RENATECH network partly supported this work for the TEM FIB thin foils preparation by David Troadec at IEMN Lille. Advice and discussions with V.A. Borodin and O. Emelyanova are gratefully acknowledged. This work has been carried out within the framework of the French Federation for Fusion FR-FCM and the Training and Education WP of the EUROfusion Consortium, funded by the European Union via the Euratom Research and Training Programme (Grant Agreement No 101052200 — EUROfusion). Views and opinions expressed are however those of the author(s) only and do not necessarily reflect those of the European Union or the European Commission. Neither the European Union nor the European Commission can be held responsible for them.

**Tables**

Table 1: Comparison between our and literature results for the lattice parameter and insertion energies of the yttria unit cell.

|  |  | $a$ (Å) | $E_{16c}^{ins}$ (eV) | $E_{8b}^{ins}$ (eV) | $E_{centre}^{ins}$ (eV) |
|---|---|---|---|---|---|
| **Our results** | **PBE** | 10.65 | 0.72 | 0.87 | 1.15 |
| **Theoretical** [24] | **GGA** | 10.69 | 0.73 | 0.87 | 1.13 |
| **Theoretical** [16] | **PBE** | 10.71 | 0.48 | 0.62 | 0.89 |
| **Theoretical** [5] | **PW91** | 10.61 | 1.22 | 0.77 | 0.92 |
| **Experimental** [106] |  | 10.60 | - | - | - |

Table 2: Our substitutional helium formation energy $E_{sub}^f$ in the yttria unit cell compared to other DFT calculations in the literature.

|  | **Type of site** | **Substitutional He energy ($E_{sub}^f$) (eV) – neutral vacancies** | **Substitutional He energy ($E_{sub}^f$) (eV) – charged vacancies** |
|---|---|---|---|
|  | Y1 | 0.383 | 0.251 |
| **Our results (PBE)** | Y2 | 0.293 | 0.295 |
|  | O | 0.822 | 0.335 |
| Other theoretical (GGA) | Y1 | 0.180 | - |

| [24] | Y2 | 0.234 | - |
| | O | 2.627 | - |

Table 3. Comparison between our pre-exponential factor and activation energies obtained with TDS and DFT for yttria. The experimental values are obtained by exponential fits weighted by the standard deviation errors of the calculated diffusion coefficients of the model. The errors are given next to the values.

| Type of sample (fluence cm$^{-2}$) | Temperature range (K/°C) | Pre-exponential factor $D_0$ (m$^2$s$^{-1}$) | Activation energy $E_a$ (eV) |
|---|---|---|---|
| **1 × 10$^{14}$ – Micrometric grains** | 453-573 / *180-300* | $5.1 \times 10^{-11} \pm 3.2 \times 10^{-11}$ | $0.871 \pm 0.029$ |
| | 573-663 / *300-390* | $7.9 \times 10^{6} \pm 3.9 \times 10^{6}$ | $2.88 \pm 0.14$ |
| **1 × 10$^{14}$ – Nanometric grains** | 483-615 / *210-342* | $3.3 \times 10^{-11} \pm 1.0 \times 10^{-11}$ | $0.777 \pm 0.015$ |
| | 615-668 / *342-395* | $1.60 \times 10^{15} \pm 0.80 \times 10^{15}$ | $4.03 \pm 0.20$ |
| **1 × 10$^{15}$ – Nanometric grains** | 493-673 / *220-400* | $2.58 \times 10^{-11} \pm 0.56 \times 10^{-11}$ | $0.730 \pm 0.052$ |
| **Interstitial diffusion (DFT)** | - | $1.25 \times 10^{-7}$ | 0.70 |

**Figures caption**

Figure 1: SRIM-calculated depth profile for helium implanted ions (appm) and the resulting damage dose (dpa) in yttria.

Figure 2: Yttria's unit bixbyite cell. The green atoms represent yttrium and the violet the oxygen atoms.

Figure 3: Visualisation of the interstitial sites in yttria. a) is the 8c and b) is the 16c site, and c) is the centre site. (I) corresponds to the visualisation of the site among all the other atoms of the yttria's unit cell. (II) is a focus of the helium atom in the site with the nearest Y and O unit cells. (III) is the He insertion site with only the immediate atoms.

Figure 4: NEB diagrams of the three possible He interstitial transitions in yttria. (a) represents the transition S1-S2 with 5 images, (b) represents the transition between S1 and S3 sites with 7 linked images and (c) represents the transition between S1-S3. (I), (II) and (III) show these transitions with the nearby atoms.

Figure 5: View of the possible diffusion pathways of helium in yttria with the respective migration energies. (I) view centered in the 8b site S1. (II) view centered in the 16c site S2. (III) view centered in the centre site S3. The dark blue arrows represent the S1-S2 transitions whereas the light blue are for the S2-S1 ones. The dark green arrows represent the S1-S3 transitions whereas the light blue are for the S3-S1 ones. The dark red arrows represent the S2-S3 transitions whereas the dark yellow are for the S3-S2 ones.

Figure 6: The transition between site S4 towards site S2. (I) shows the transition with the nearest yttrium and oxygen sites. (II) is the pathway of eleven linked system images between the sites S4 to S2 and the energy comparison between the images and the sites. The colour light green represents the S4 site, orange for the S2, darker green for the images, and purple for the saddle point (transition state). The big green spheres represent the yttrium atoms, and the pink ones for the oxygen atoms.

Figure 7: Transmission electron microscopy bright-field images showing the lack of visible helium bubbles in the yttria micrograined samples. (I) and (II) are underfocused and overfocused images of ±1.0 µm centred at the $R_p$ (~300 nm depth), and (a) and (b) are underfocused and overfocused images of ±0.5 µm at the grain boundary, also located at a depth around the projected range $R_p$, respectively.

Figure 8: a) Cumulated helium desorption profiles from the TDS versus the temperature for a heating ramp of 5°C/min and for pure yttria with nanometric grains. b) Instantaneous release as a function of the temperature for pure yttria with nanometric grains. c) Cumulate helium release profiles of nanometric and micrometric yttria grains from the TDS versus the temperature for a heating ramp of 5 °C/min. d) Instantaneous release as a function of the temperature. The different peak regions are indicated in the figure.

Figure 9: Cumulated helium cumulated release fraction from TDS versus time for three different samples for plateaux experiments. a) pure yttria with micrometric grains implanted with He under $1 \times 10^{14}$ at.cm$^{-2}$. b) pure yttria with nanometric grains implanted with a He fluence of $1 \times 10^{14}$ at. cm$^{-2}$. c) pure yttria with nanometric grains implanted with He at $1 \times 10^{15}$ at. cm$^{-2}$. The temperature profile is shown on the right side. In a), the model results and the relative error between experimental and simulated results are shown.

Figure 10: Cumulated and instantaneous He release profile from TDS versus time from the experimental data (blue) and the simulated curve (orange) for the yttria $1 \times 10^{14}$ with micrometric grains sample experiment with a 5 °C ramp. a) Cumulated He release profile as a function of the time. b) Instantaneous release as a function of time.

Figure 11: Comparison of our DFT and TDS calculated He diffusion coefficients in pure yttria versus the reciprocal temperature (400 to 1100 K). The exponential fits are done by weighted standard deviations of the diffusion coefficient values calculated by the model.

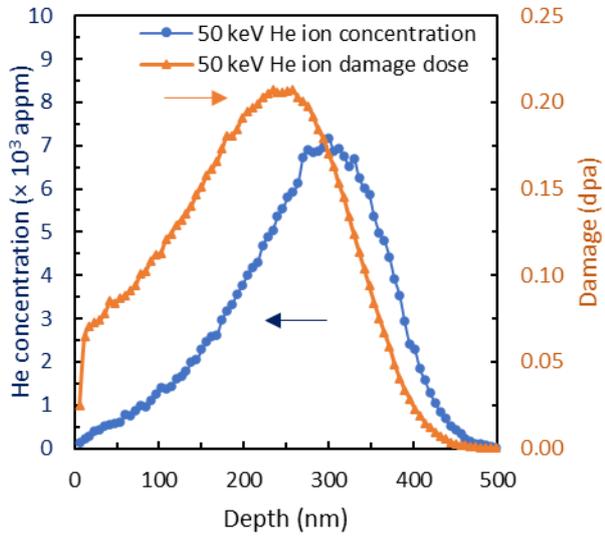

Figure 1

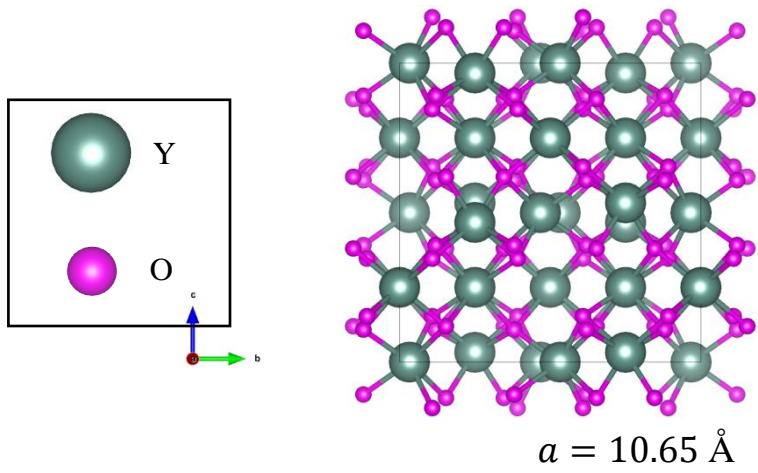

$a = 10.65$ Å

Figure 2

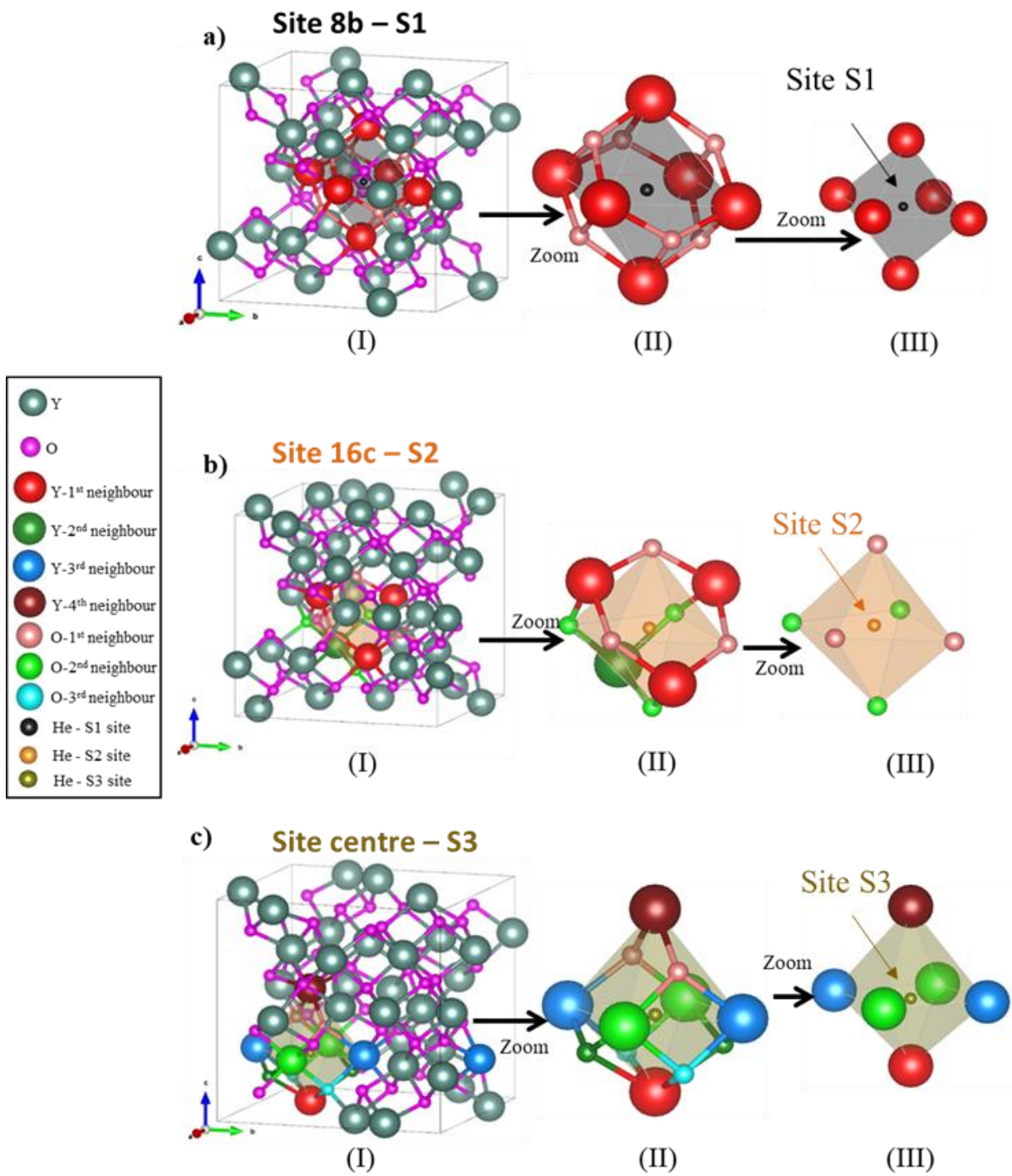

Figure 3

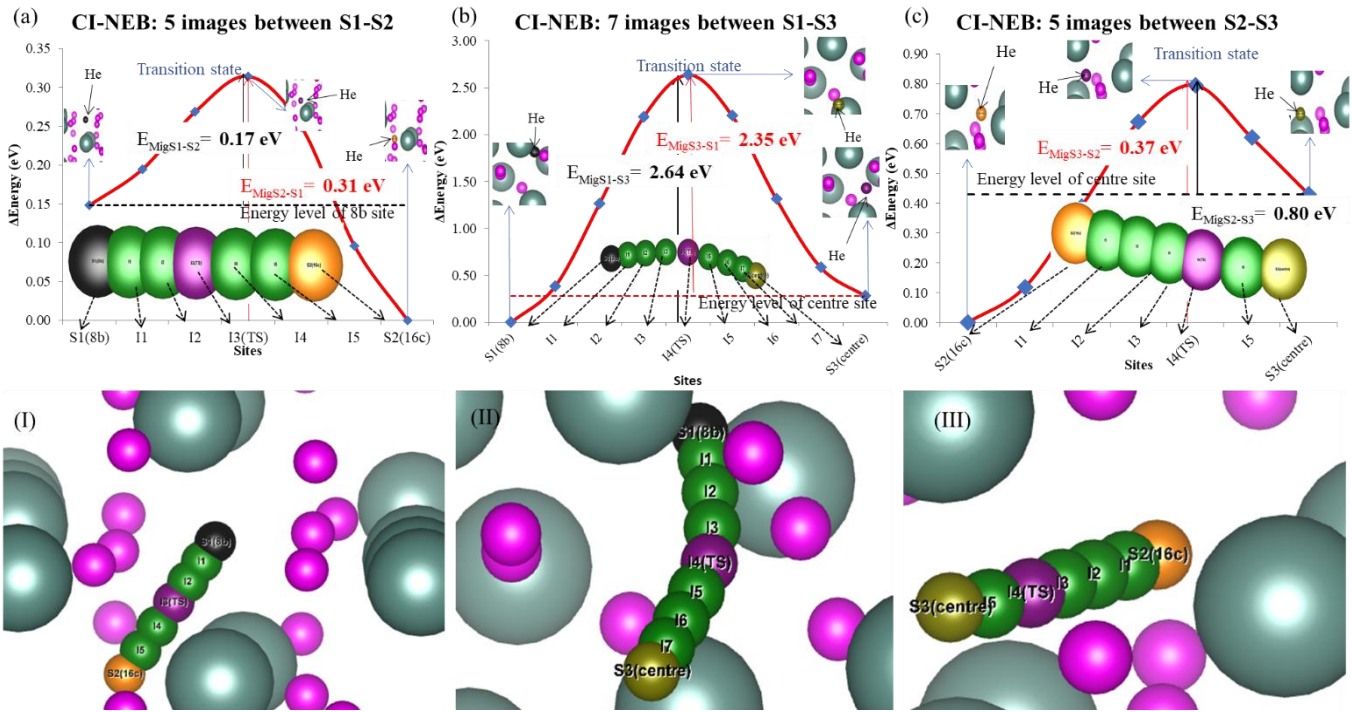

Figure 4

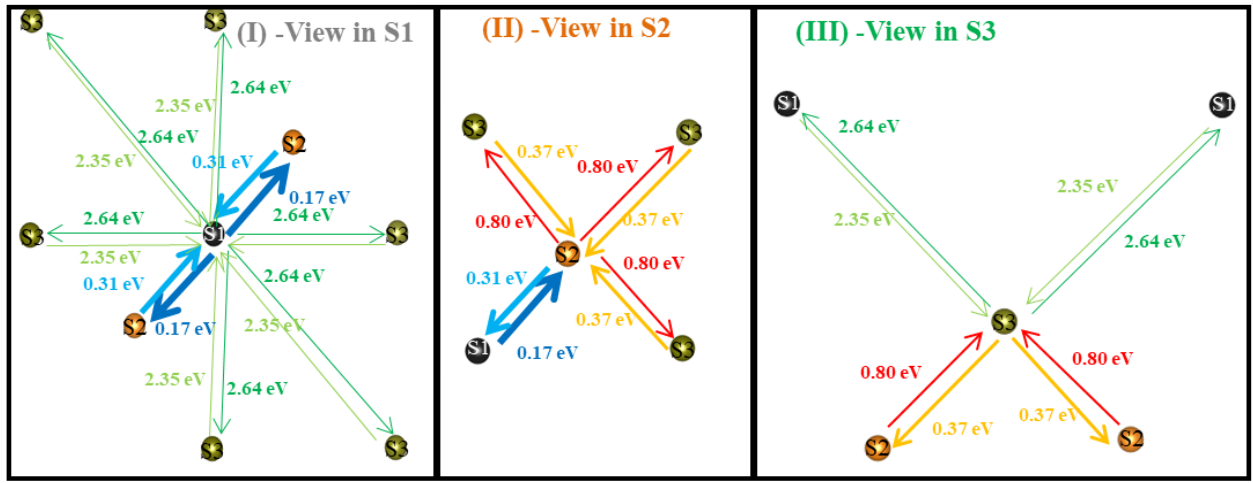

Figure 5

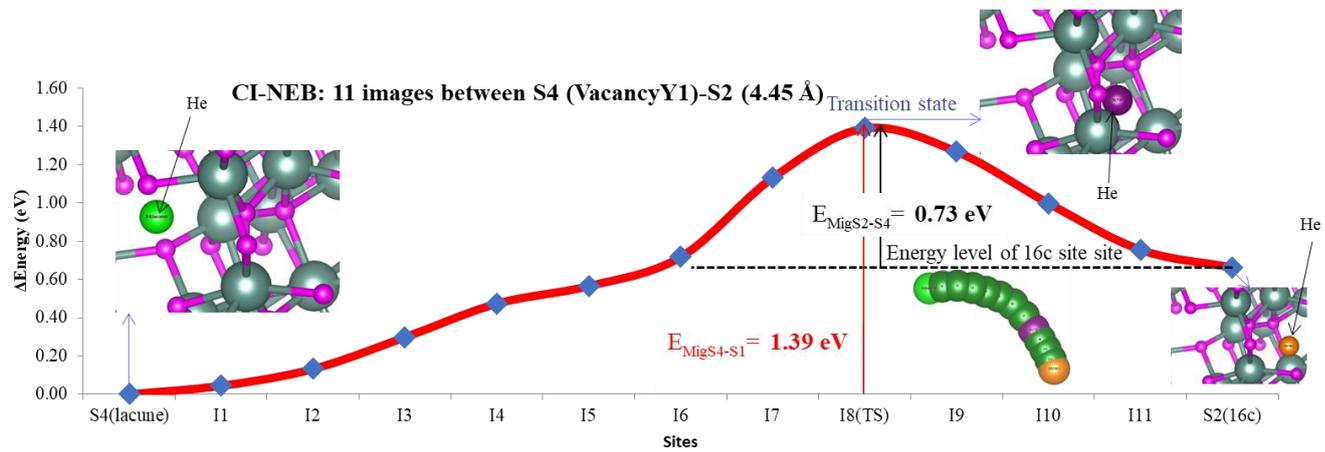

Figure 6

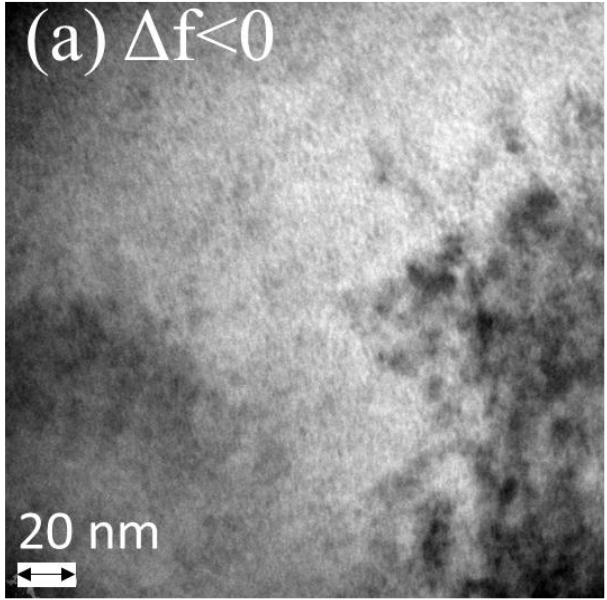 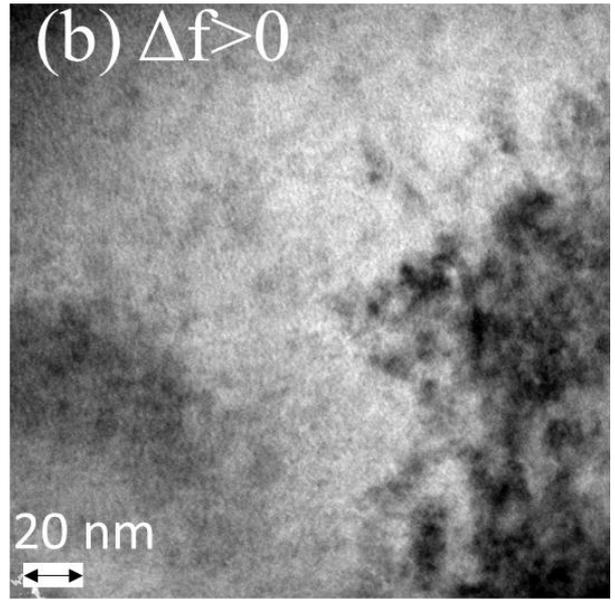
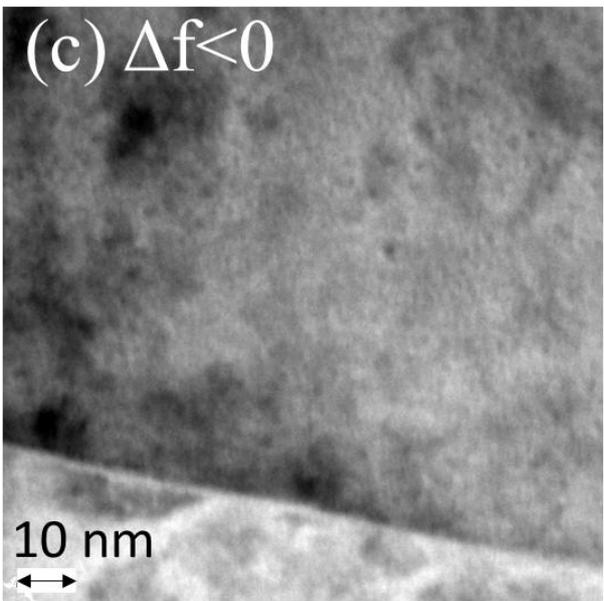 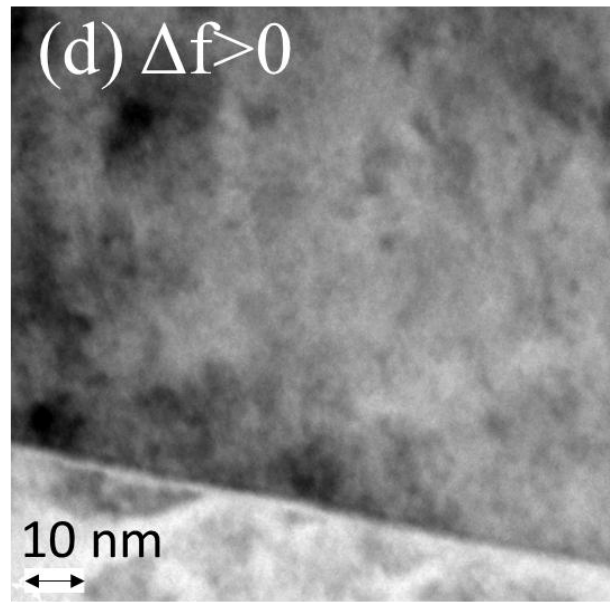

Figure 7

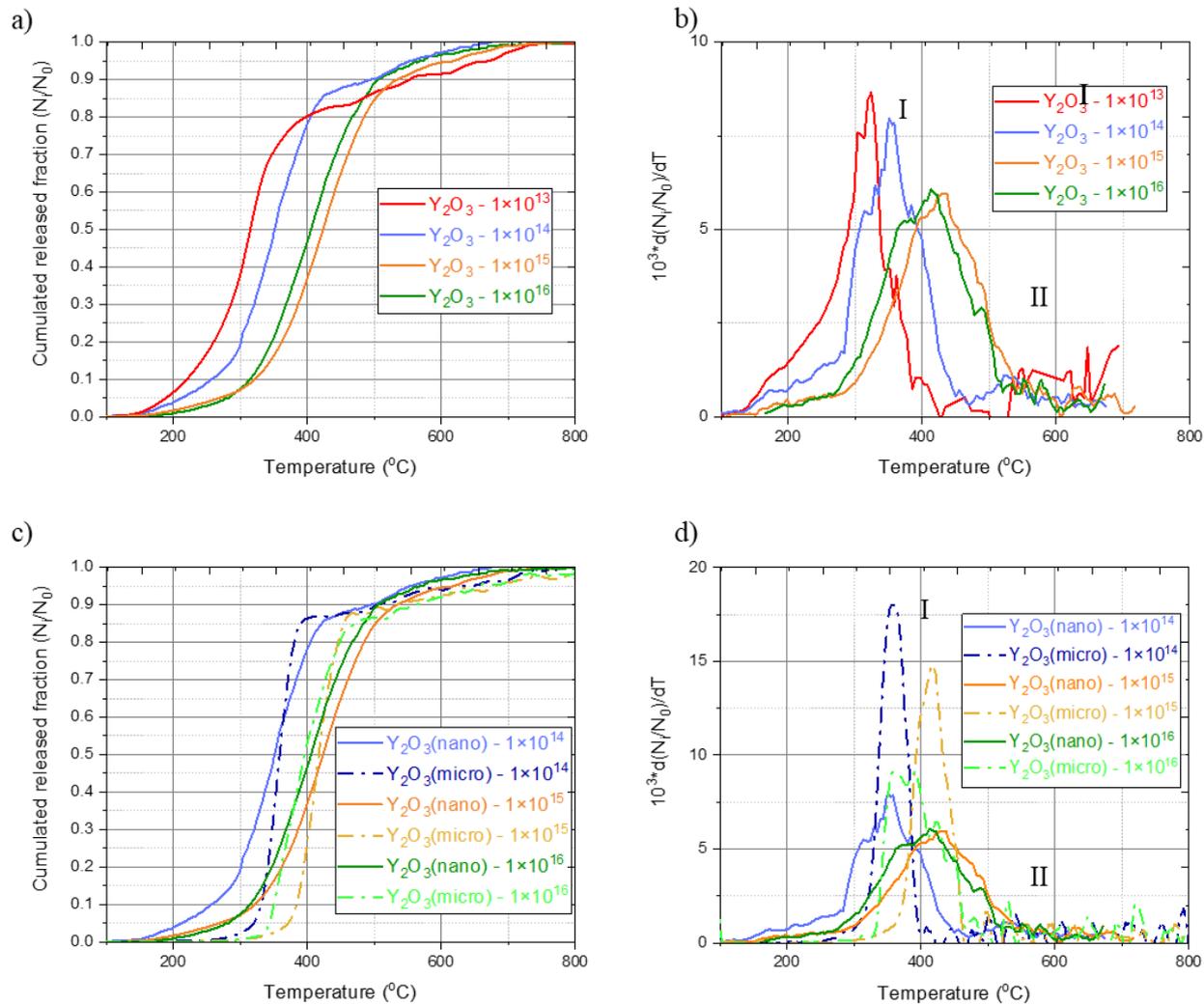

Figure 8

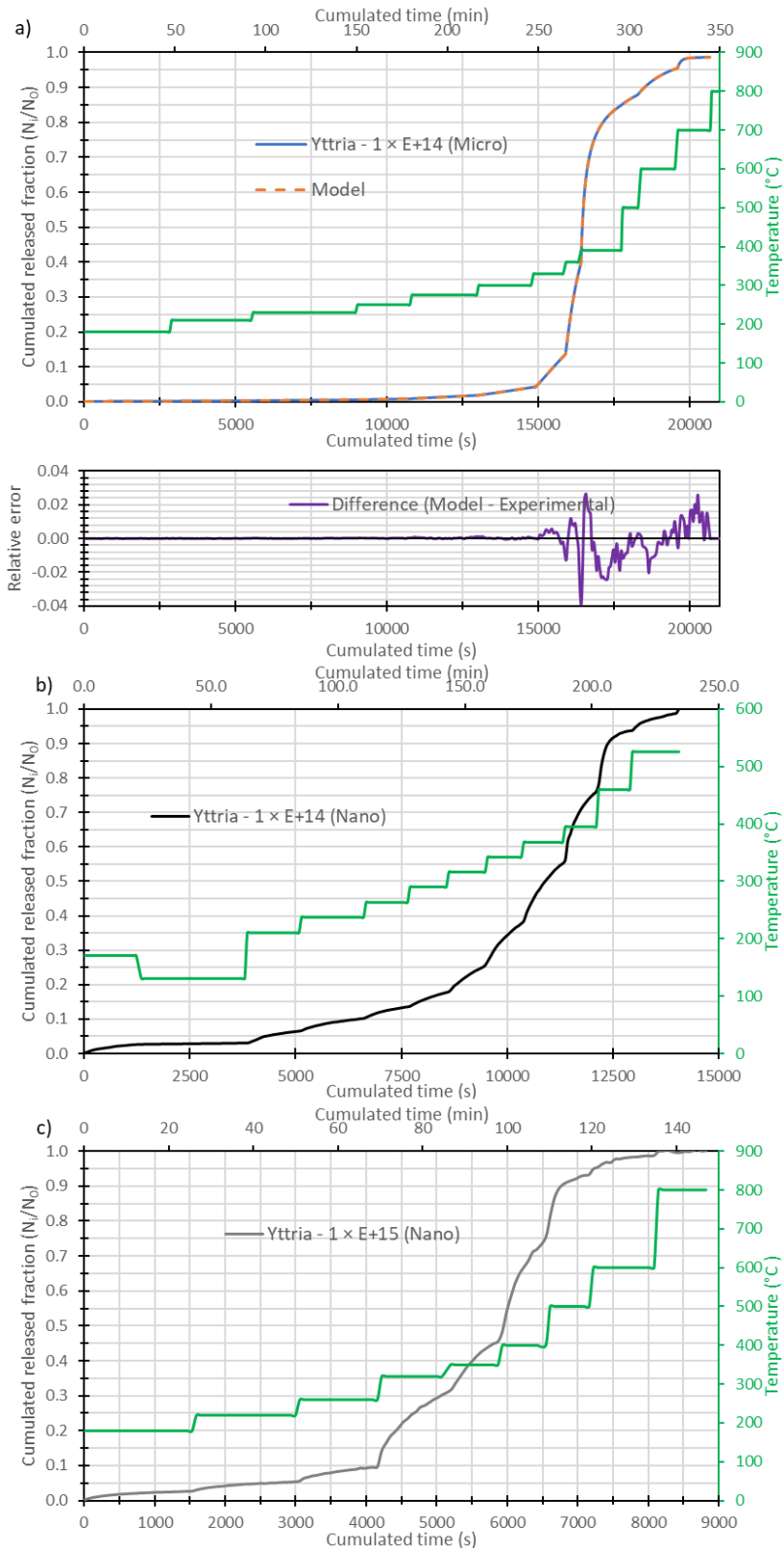

Figure 9

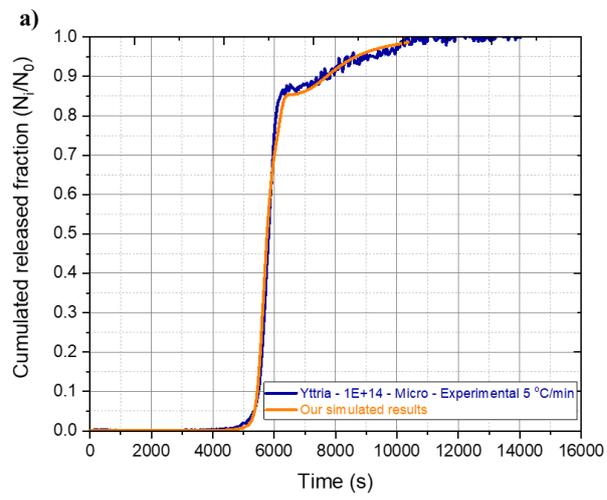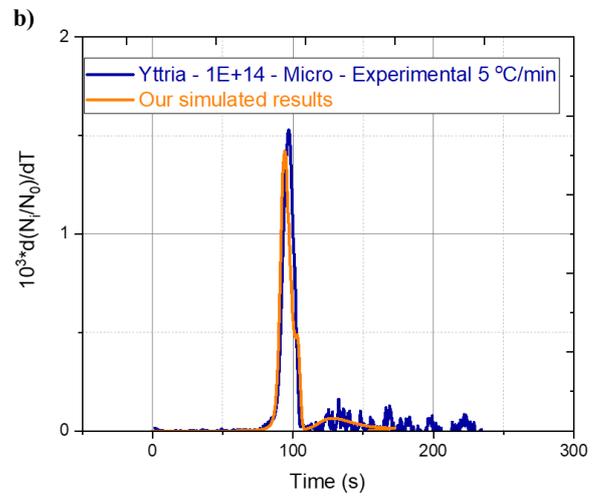

Figure 10

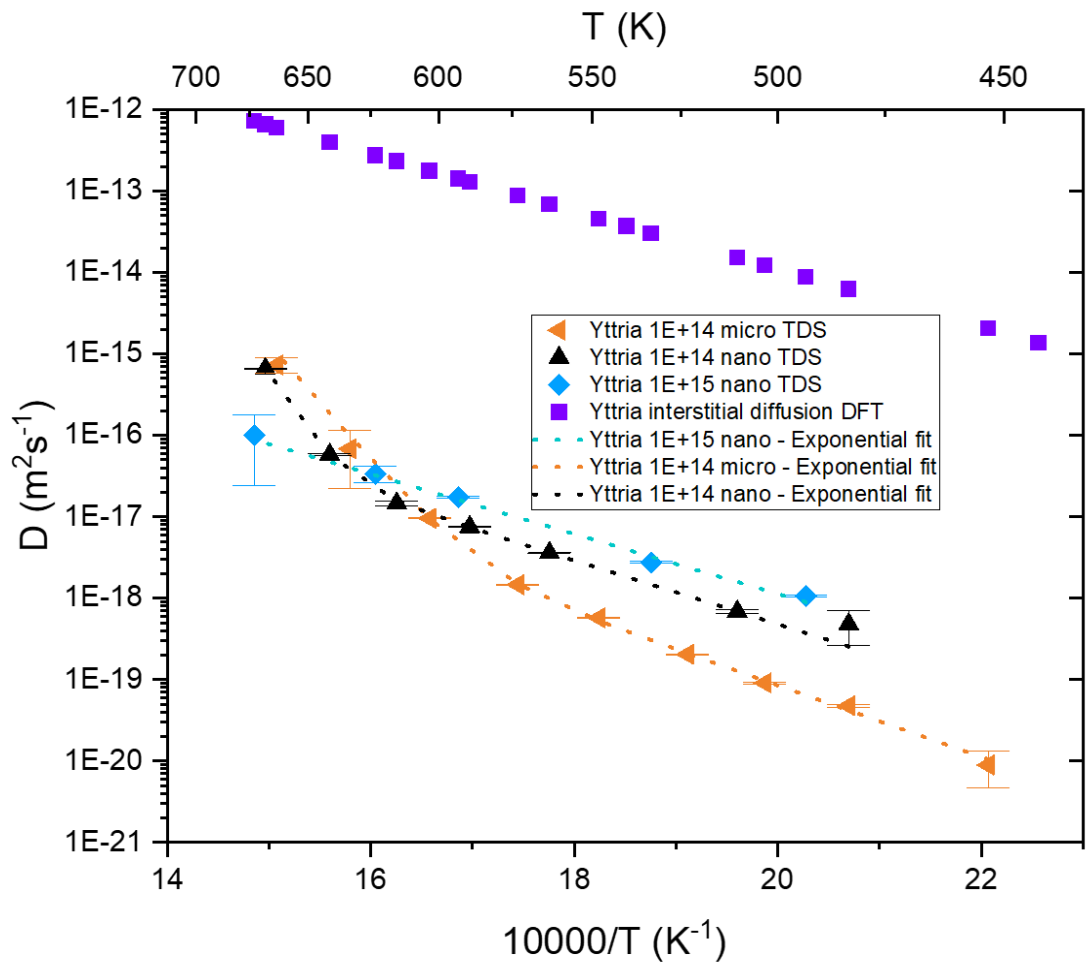

Figure 11